\documentclass[12pt,a4paper,final]{iopart}

\usepackage{iopams}
\usepackage{cite}

\expandafter\let\csname equation*\endcsname\relax

\expandafter\let\csname endequation*\endcsname\relax

\usepackage{amsmath}
\usepackage{iopams}
\usepackage{graphicx}
\usepackage[breaklinks=true,colorlinks=true,linkcolor=blue,urlcolor=blue,citecolor=blue]{hyperref}

\usepackage{soul}

\begin{document}
	
	\title[Survival in two-species reaction-superdiffusion system]{Survival in two-species reaction-superdiffusion system: Renormalization group treatment and numerical simulations}	
	
	\author{Dmytro Shapoval$^{1,2}$, Viktoria Blavatska$^{1,2,3}$, Maxym Dudka$^{1,2}$}
	\address{$^1$Institute for Condensed Matter Physics, National Academy of Sciences of Ukraine, 1 Svientsitskii Street, UA-79011 Lviv, Ukraine}
	\address{$^2$ ${\mathbb L}^4$ Collaboration \& Doctoral College for the Statistical Physics of Complex Systems, Leipzig-Lorraine-Lviv-Coventry, Europe}
	\address{$^3$Dioscuri Centre for Physics and Chemistry of Bacteria,	Institute of Physical Chemistry, Polish Academy of Sciences, 01-224 Warsaw, Poland}
	
	\begin{abstract}
		We analyze the two-species reaction-diffusion system including trapping reaction $A + B \to A$ as well as coagulation/annihilation reactions $A + A \to (A,0)$  where  particles of  both species are performing  L\'evy flights with control parameter $0 < \sigma < 2$, known to lead to superdiffusive behaviour. The density  as well as the correlation function   for target particles $B$  in such systems are known to scale  with nontrivial universal exponents at space dimension $d \leq d_{c}$. Applying the renormalization group formalism we calculate these  exponents in a case of superdiffusion below the critical dimension  $d_c=\sigma$.  The numerical simulations in one-dimensional case are performed as well. The quantitative estimates for the decay exponent of the density of survived particles  $B$  are in a good agreement with our analytical results. In particular, it is found that the surviving probability of the target particles in superdiffusive regime  is higher than that in a system with ordinary diffusion.
	\end{abstract}
	
	
	\maketitle

\section{Introduction}
\setcounter{footnote}{0}

Reaction-diffusion models are exploited to describe  the peculiarities  of a wide range of non-equilibrium systems of many interacting agents in such various fields as physics, chemistry, biology, ecology, sociology, and economics \cite{kotomin1996, henkel2008, odor2008, kraprivsky2010,tauber2014}. Of special interest  are the systems possessing reactions of trapping type $A + B \to A$, which are  widespread in nature: one should mention  predator-pray ecological systems \cite{predator}, binding of proteins systems \cite{binding}, quenching of localized excitations \cite{excitation} as examples. 


If the target particles $B$ are immobile in the medium containing diffusive traps $A$, such a problem is called {\em target problem} in the literature. Reciprocal problem, when traps $A$ are static and only target particles $B$ diffuse, is known as {\em trapping problem} (see e. g. reviews \cite{problem}). While the former problem possesses exact solution for survival probability of target particles within the frames of Smoluchowski approximation \cite{Smoluch_appr} (for original Smoluchowski' work see \cite{Smoluchowski}), exact results for the latter case can be obtained only in limiting cases (see e. g. review in \cite{Yuste08}). 

The case when both particles $A$ and $B$ are allowed to diffuse is more interesting and is closer to reality. Studies  of asymptotic decay of survival probability demonstrate nontrivial correlations between target particles and traps for spatial dimensions $d\leq 2$, which invalidate the rate equation description \cite{BramsonLebowitz}. The space dimension $d_c=2$ is known to be the critical one for the large variety of reaction-diffusion systems with irreversible reactions, below which evolution on a long time scale is diffusion-controlled (see e. g. \cite{tauber2014}). 

Impact of fluctuations in  reaction-diffusion systems for $d \leq d_c$ has been explored using several methods, in particular: Smoluchowski-type approximations \cite{krapivsky1994,Ovchinnikov89}, many-particles density formalism \cite{Kotomin92}, weakly non-ideal Bose gas approximation \cite{Ovchinnikov89, Konkoli04} as well as  field-theoretic renormalization group (RG) approach \cite{tauber2014,tauber2005,Hnatic2013}. Among them, the RG  technique offers powerful methods to analyse   systematically the large-time asymptotic behaviour of reaction-diffusion models.

In the present paper, we apply RG methods to  analyze  the problem of  particles' survival  in a media with mobile traps in a diffusion-controlled  regime, represented by  two-species reaction-diffusion system:
\begin{eqnarray}
	\label{model}
	A + A &\to& 
	\begin{cases}
		A, \, \, \text{with probability} \, p \,  \text{(coalescence)},\\
		0, \, \, \text{with probability} \, 1-p \,  \text{(annihilation)},
	\end{cases} \nonumber \\
	A + B &\to& A \, \, (\text{trapping}),
\end{eqnarray}
where target particles ($B$) are absorbed by traps ($A$ particles) and the trap particles may coagulate or mutually annihilate. ``Target problem'' (when $B$ particles are static) in this case for $d=1$ is equivalent to the problem of estimation of amount of spins not flipped up to time $t$ in the $q$-state Potts model with zero temperature Glauber dynamics, where $q=1-1/(1-p)$ \cite{derrida1995}. When particles of both species  can diffuse, decay  of $B$ particle density  for the case $p=1$ in $d=1$ is equivalent to survival probability of the three vicious walker problem  \cite{fisher1988,Redner01}. RG studies have shown that $B$ particle density decays    \cite{rajesh2004,vollmayrlee2018} with nontrivial universal exponent $\theta$, which is dependent on dimension $d$, trap reaction parameter $p$, and the ratio of diffusion constants $\delta$. Recently, it was demonstrated  by RG approach that correlation function  of $B$ particles also has scaling behavior with universal exponent $\phi$, found to be dependent on $d$ and $p$ in the first nontrivial approximation \cite{vollmayrlee2018}. Numerical calculations for cases $d=1$ and $d=2$ corroborate RG theory outcome for model (\ref{model}) \cite{vollmayrlee2020}.

Although ordinary diffusion is usually  considered in non-equilibrium models, the complex structure  of some systems may lead to the non-Markovian behaviour (see e. g. \cite{Olivera19}), violating the law of linear growth of mean square displacement   with time $t$, so that 
\begin{equation}\label{msd}
	\langle r^2\rangle\sim t^\alpha
\end{equation}
with $\alpha\not=1$ (while $\alpha=1$ for normal diffusion). Systems with subdiffusive motion ($0<\alpha<1$) of traps and/or of target particles are intensively considered in the literature (see Ref. \cite{Yuste08}). Less attention was paid to study the trapping reactions in system with anomalous superdiffusion.  The latter can be modeled exploting the idea of  L\'evy flights represented by random walks with discrete jump lengths obeying the probability distribution with long tails. In each time step, the particles' jump of length $r$ is chosen from the L\'evy distribution 
	\begin{equation} 
		{\hat p}_L(r) \sim r^{- d - \sigma},\label{dis}
	\end{equation} 
with the control parameter $0 < \sigma \leqslant 2$, and the spatial dimension $d$ {\footnote{  As discussed in the literature, the L\'evy  flight statistics  may lead to divergent mean square displacement (an infinite velocity), which is unphysical. In contrast, in model of anomalous diffusion called L\'evy walk, each the discrete time to make a jump of some size is taken proportional to this size and particles have a finite velocity. (for more details see \cite{dybiec2017, palyulin2019, benichou2011}).}}. In this case, the mean square displacement (\ref{msd}) grows with $\alpha=2/\sigma$. 

L\'evy flights have been widely applied in description of  many processes realized in nature \cite{review__levy_app__1, review__levy_app__2} including nonlinear dynamics, in particular, chaotic diffusion in Josephson junctions \cite{josephson_junctions}, streamline transport properties \cite{chernikov1990}, the dynamics of particles in periodic potentials \cite{chaikovsky1991} or turbulent flow \cite{turbulent_flow}. Furthermore, such superdiffusive random process have been used to describe the biological systems, in particular anomalous diffusion in living polymers \cite{ott1990}, as well as the diffusion of DNA-binding proteins \cite{lomholt2009},  long-range spreading of epigenetic marks along the genome \cite{Ancona2020}. L\'evy flights have been used to describe laser cooling of cold atoms \cite{bardou1994} and the spectral fluctuations in random lasers \cite{random_lasers} or light transport in optical materials with impurities (so-called L\'evy glasses) \cite{levy_glasses}, in  optimization the search strategies \cite{search_strategies}. In analogy with usual L\'evy flights the  ``temporal'' L\'evy flights were introduced, with motion of  particles characterized by algebraically distributed waiting times. Such temporal L\'evy flights can lead to subdiffusion \cite{Hinrichsen07,Stanislavsky08}. We will consider a model (\ref{model}) with the anomalous diffusion, which is modeled by super-diffusing L\'evy flying particles, i.e., in each time step particles' movement is chosen from the L\'evy distribution (\ref{dis}). 

Field-theoretical renormalization group technique has been applied in a  variety of reaction-diffusion models  with L\'evy flights including single-species annihilation reaction \cite{vernon2003}, branching and annihilating models \cite{vernon2001}, anomalous directed percolation \cite{hinrichsen1999,Hinrichsen07,janssen2008}, vicious walkers \cite{Goncharenko10} and demonstrated that in such cases  the upper critical dimension is determined by the control parameter for the L\'evy distribution, $\sigma$.

The first attempts to study the model (\ref{model}) with Lev\'y flights in a regime far  below the critical dimension, which is $d_{c} = \sigma$,  have been performed in Ref. \cite{birnsteinova2020}. The case   $\sigma < 2$ was considered, that allowed  to neglect terms related to the ordinary diffusion, that appeared in a field-theoretical representation. In our paper, we exploit the same limit and calculate the universal exponents characterising long-time asymptotic scaling behavior of density and correlation function for $B$ particles.    

The paper is outlined as follows: in Sec.~\ref{field_theory} we introduce the field theory for our model with constituent elements of the corresponding Feynman diagrams. In Sec.~\ref{renormalization} we describe application of RG to find exponents governing long-time  behavior of density and correlation function for $B$ particles. In Sec.~\ref{results} we present results for these quantities. In Sec.~\ref{V} we present results of numerical simulations of our model on the one-dimensional lattice and compare them with the theoretical estimates. Finally, we summarize our study by conclusions in Sec.~\ref{conclusions}. { In the Appendix~\ref{ap0} we briefly describe Smoluchowski approximation for our model, while details concerning the one-loop contribution to analyzed quantities are presented in the Appendix~\ref{apI}.
	
\section{Field-theoretical description}
\label{field_theory}
	
In order to study long-time scaling behavior of observables in reaction-diffusion models,  it is standard now to appeal to well-grounded RG methods \cite{tauber2014,tauber2005,Hnatic2013}. It is applied for the effective action obtained within the field theory representation.  Using standard technique \cite{tauber2005, doi1976, peliti1985}, this representation can be obtained mapping master equations to effective theory. Master equation for our model is presented in following subsection.

\subsection{Master equation}
\setcounter{footnote}{1}
	
Let us  consider that our traps ($A$ particles) and targets ($B$ particles) may  occupy  sites of a $d$-dimensional hypercubic lattice.  The  particles of both species hop from one site to another due to the L\'evy distribution (\ref{dis}). At a rate $\lambda$ two $A$ particles at the same site may annihilate each other or coagulate together, reducing the number of $A$ particles, moreover, an $A$ particle may absorb $B$ particle meeting it at the same site at a rate $\lambda^{\prime}$ (\ref{model}). 
	
We describe such two-species reaction-diffusion model in terms of the master equation  for temporal  probability distribution  $P(\{n^A\};\{n^B\})$ of a configuration characterized by sets of occupation numbers $\{n^A\}=\dots,n^A_i\dots$ for traps $A$ and $\{n^B\}=\dots,n^B_i\dots$ for target particles $B$, $i=1,\dots,N$, with $\sum_i n^A_i=N_A(t)$ and $\sum_i n^B_i=N_B(t)$. Change of temporal  probability distribution can be decomposed into contributions coming  from L\'evy flights transport, coagulation, annihilation and trapping reactions separately:
	\begin{flalign}\label{mse}
		\frac{d P(\{n^A\};\{n^B\})}{d t}&{=}\left.\frac{\partial P(\{n^A\};\{n^B\})}{\partial t}\right|_L\! +\left.\frac{\partial P(\{n^A\};\{n^B\})}{\partial t}\right|_A +\left.\frac{\partial P(\{n^A\};\{n^B\})}{\partial t}\right|_C&& \nonumber\\
		&+\left.\frac{\partial P(\{n^A\};\{n^B\})}{\partial t}\right|_T.&&
	\end{flalign}
The first term in the r.h.s. terms  of Eq. (\ref{mse})  is a change of the probability distribution ${P(\{n^A\}{,}\{n^B\})}$ at jumps of the particles between sites, while  the second and third terms  present a change of $P(\{n^A\},\{n^B\})$ at the annihilation and coagulation reactions, the last one gives a change of $P(\{n^A\},\{n^B\})$ at the trapping reactions.	Master equation for L\'evy jumps of particles between sites $j$ and  $i$    can be written in the following form 
	\begin{flalign}\label{levy}
		\left.\frac{\partial P(\{n^A\};\{n^B\})}{\partial t}\right|_L&=\sum_{i\not{=}j}{\hat p}_L(r_{ij})D_{0A}\Big [(n^A_j+1)P({\dots},n^A_i-1,{\dots},n^A_j+1,{\dots};\{n^B\})&&\nonumber\\
		&-n^A_jP(\{n^A\};\{n^B\})\Big]&&\nonumber\\
		&{+}\sum_{i\not{=}j}{\hat p}_L(r_{ij})D_{0B} \Big[(n^B_j+1)P(\{n^A\};{\dots},n^B_i-1,{\dots},n^B_j+1,{\dots})&&\nonumber\\
		&-n^B_jP(\{n^A\};\{n^B\})\Big],&&
	\end{flalign} 
with ${\hat p}_L(r)$ given by Eq. (\ref{dis}), $D_{0A}$ and $D_{0B}$ are transport (diffusion) constants for traps and target particles respectively, summation is over all pairs $i$ and $j$.	Annihilation and a coagulation reactions is described by following master equations
	\begin{eqnarray}
		\left.\frac{\partial P(\{n^A\};\{n^B\})}{\partial t}\right|_A=& \frac{\lambda_{A}}{l_0^d}&\sum_i\Big [(n^A_i+2)(n^A_i+1)P({\dots},n^A_i+2,{\dots};\{n^B\})\nonumber\\&-&n^A_i (n^A_i-1) P(\{n^A\};\{n^B\})\Big],\\
		\left.\frac{\partial P(\{n^A\};\{n^B\})}{\partial t}\right|_C=& \frac{\lambda_{C}}{l_0^d}&\sum_i\Big [(n^A_i+1)n^A_i P({\dots},n^A_i+1,{\dots};\{n^B\})\nonumber\\&-&n^A_i(n^A_i-1)P(\{n^A\};\{n^B\})\Big],
	\end{eqnarray} 
where annihilation reaction rate is $\lambda_A=(1-p)\lambda$ and coagulation reaction rate $\lambda_C=p\lambda$. Last term in (\ref{mse}) concerning trapping reaction obeys following master equation:
	\begin{eqnarray}\label{trap}
		\left.\frac{\partial P(\{n^A\};\{n^B\})}{\partial t}\right|_T=&\frac{\lambda'}{l_0^d}&\sum_i\Big [n^A_i(n^B_i+1)P({\dots},n^A_i,{\dots};{\dots},n^B_i+1,{\dots})\nonumber\\&-&n^A_i n^B_i P(\{n^A\};\{n^B\})\Big]
	\end{eqnarray} 
	
Master equation  (\ref{mse}) with (\ref{levy})-(\ref{trap}) describes the microscopic behavior of our  system and can be  mapped to a field theory  using standard technique \cite{tauber2005, doi1976, peliti1985}. We present this theory in next section.
	
\subsection{Field theory}
	
Mapping of the master equation to the field theory is performed within  standard technique representing it in terms of the second-quantized bosonic operators in Fock space and then constructing path-integral Doi-Peliti representation  for coherent state basis of resulting non-Hermitian problem \cite{doi1976, peliti1985} (see also \cite{tauber2014,tauber2005}).  While transformation of term (\ref{levy}) for ordinary diffusion case ($|r_{ij}|=1$) leads to the action for model with terms  $D_{oA}\nabla^2$ and $D_{oB}\nabla^2$, where diffusion constants  $D_{oA}$ and $D_{oB}$ are related to $D_{0A}$ and $D_{0B}$ correspondingly, the field theory action of  our model with L\'evy flights instead such terms  has
	\begin{equation}
		D_{oA}\nabla^2\to D_{oA}\nabla^2+D_{A}\nabla^\sigma,\quad D_{oB}\nabla^2\to D_{oB}\nabla^2+D_{B}\nabla^\sigma,
	\end{equation}
for details see e.g. \cite{hinrichsen1999}. $\nabla^\sigma$ is a symbolic notation of the operator defined by its action in momentum space \cite{vernon2003}:
	\begin{equation}
		\nabla^\sigma e^{i {\bf k}{\bf r}} =-|k|^\sigma e^{i{\bf k}{\bf r}}.
	\end{equation}
As was already noted \cite{vernon2003,birnsteinova2020},  the normal diffusion terms $~\sim \nabla^2$  are irrelevant for the case $\sigma<2$  and therefore can be dropped out from the field theoretical description. Note, however, to describe the behaviour near $\sigma\to 2$, both terms (for normal diffusion and for anomalous diffusion) must be taking into account.
	
As a  final result of Doi-Peliti procedure  an effective field theory describing behavior of the two-species reaction-diffusion model (\ref{model}) with L\'evy flights for its action  we use \cite{birnsteinova2020}:
	\begin{eqnarray}
		\label{action}	
		S &=& \int d^{d} x \, d t \left\{ \bar{a} \left(\partial_{t} - \nabla^{\sigma}\right)a + \bar{b} \left(\partial_{t} - \delta\nabla^{\sigma}\right)b \right. \nonumber \\
		&+& {\lambda} \bar{a} a^{2} + \lambda \bar{a}^{2} a^{2} + \lambda^{\prime} Q\bar{b} a b + \lambda^{\prime} \bar{a} \bar{b} a b + \left.\left(\bar{a} a_{0} + \bar{b} b_{0}\right) \delta\left(t\right) \right\},
	\end{eqnarray}
where  $\delta = D_{A}/D_{B}$, $a$ and $b$ are complex fields corresponding to $A$ and $B$ particles, while response fields $\bar{a}$ and $\bar{b}$ play the same role as auxiliary fields in Martin-Siggia-Rose approach for critical dynamics \cite{tauber2014}. 
The first line of (\ref{action}) describes an anomalous diffusion movement of  particles, while the second one (without the last term) corresponds to the  reactions. Value $Q = 1 / (2 - p)$  appears first at annihilation/coalescence term $\bar{a} a^{2}$ since coagulation and annihilation contribute to it as $2\lambda_A+\lambda_C=(2-p)\lambda$,  after proper rescaling ($a \to Q a$, $\bar{a} \to \bar{a} / Q$,   $a_0 \to Q a_0$ \cite{rajesh2004, vollmayrlee2018}) we finally got it at  trapping reaction term $\bar{b} a b$. Last term of (\ref{action}) corresponds to Poissonian initial conditions at $t = 0$ with an average densities $a_{0}$ and $b_{0}$. Simple power counting on (\ref{action}) reveals that the upper critical dimension, below which fluctuations effects become important, is $d_c=\sigma$.
	
The averages of an observable $O$ within the field theory with action (\ref{action}) can be calculated via functional integral
	\begin{equation}
		\langle O(t)\rangle ={\cal N}^{-1}\int{\cal D}[a,b,\bar{a},\bar{b}] O(a(t),b(t)) e^{-S}, \quad {\cal N}=\int{\cal D}[a,b,\bar{a},\bar{b}] e^{-S}.
	\end{equation}
We are interested in the mean density of target particles $\langle b(t) \rangle$ as well as in the correlation function $\tilde{C}_{BB}(r,t)=\frac{\langle b(r,t)b(0,t) \rangle-\langle b(t) \rangle^2}{\langle b(t) \rangle^2}$. We consider that density of $B$ particles for large $t$ scales as 
	\begin{equation}\label{scal1}
		\langle b(t)\rangle \sim t^{-\theta}.
	\end{equation}
while correlation function scales as
	\begin{equation}\label{scal2}
		\tilde{C}_{BB}(r,t) \sim t^{\phi}f(r/t^{1/\sigma}).
	\end{equation}
	
Calculating observables by standard methods of perturbation theory we develop  perturbative expansions in powers of coupling constants $\lambda$, $\lambda'$ and express them in form of Feynman diagrams using building element listed in Fig.\ref{propagators_and_vertices}. The standard way is to group diagrams for each calculated quantity according to number of loops. There is infinitive number of diagrams at each loop level. There the tree-level (without loop) diagrams are of importance to get mean-field densities and dressed  propagators. The infinite sums of tree level diagrams for  densities $\langle a(t) \rangle_{tr}$, $\langle b(t) \rangle_{tr}$ are obtained from Dyson equations (see Fig.~\ref{dyson_eqs} for graphical representation) resulting  in
	\begin{eqnarray}
		\label{tree-level_densities}
		\langle a(t) \rangle_{tr} &=& \frac{a_{0}}{1 + \lambda a_{0} t}, \nonumber \\
		\langle b(t) \rangle_{tr} &=& \frac{b_{0}}{(1 + \lambda a_{0} t)^{Q \lambda^{\prime}/ \lambda}}.
	\end{eqnarray}
In turn, the dressed propagators obtained via Dyson equations (see Fig.~\ref{dyson_eqs} for graphical representation)  read as:
	\begin{eqnarray}
		\label{dressed_propagators}
		G_{tr}^{AA} \left(\vec{k}, t_{2}, t_{1}\right) &=& 	\Theta(t_2-t_1)\left( \frac{1 + \lambda a_{0} t_{1}}{1 + \lambda a_{0} t_{2}}\right)^{2} e^{-k^{\sigma}(t_{2} - t_{1})};
		\nonumber \\
		G_{tr}^{BB} \left(\vec{k}, t_{2}, t_{1}\right) &=& 
		\Theta(t_2-t_1)\left( \frac{1 + \lambda a_{0} t_{1}}{1 + \lambda a_{0} t_{2}}\right)^{Q \lambda^{\prime}/ \lambda} e^{-\delta k^{\sigma}(t_{2} - t_{1})},
	\end{eqnarray}
where $\Theta(x)$ is the Heaviside step  function.
	
Tree level diagrams for  $\tilde{C}_{BB}$ are shown in Fig.~\ref{tree_corr}
	
	\begin{figure}[t!]
		\begin{center}
			\includegraphics[width=90mm]{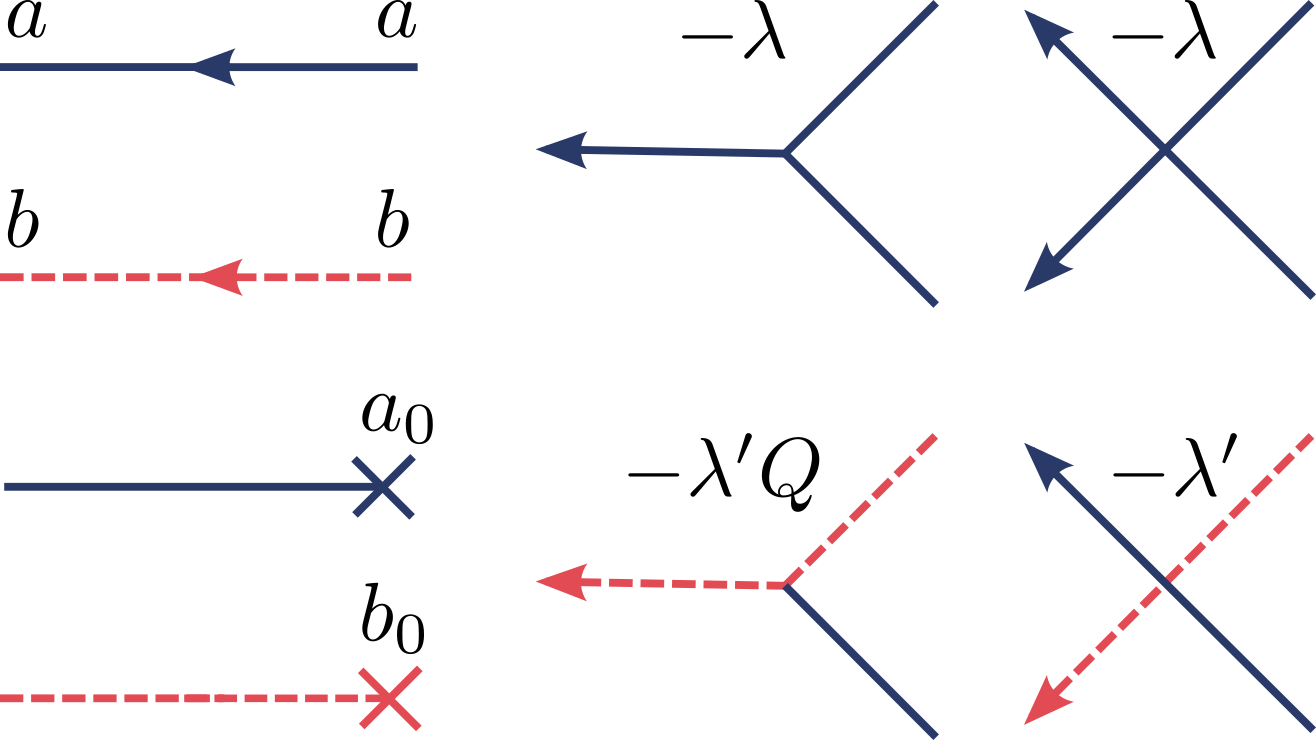}
			\caption{Building blocks for Feynman digramms of theory with action (\protect\ref{action}): propagators (left column) and vertices.}
			\label{propagators_and_vertices}	
		\end{center}
	\end{figure}
	
	\begin{figure}
		\begin{center}
			\includegraphics[width=80mm]{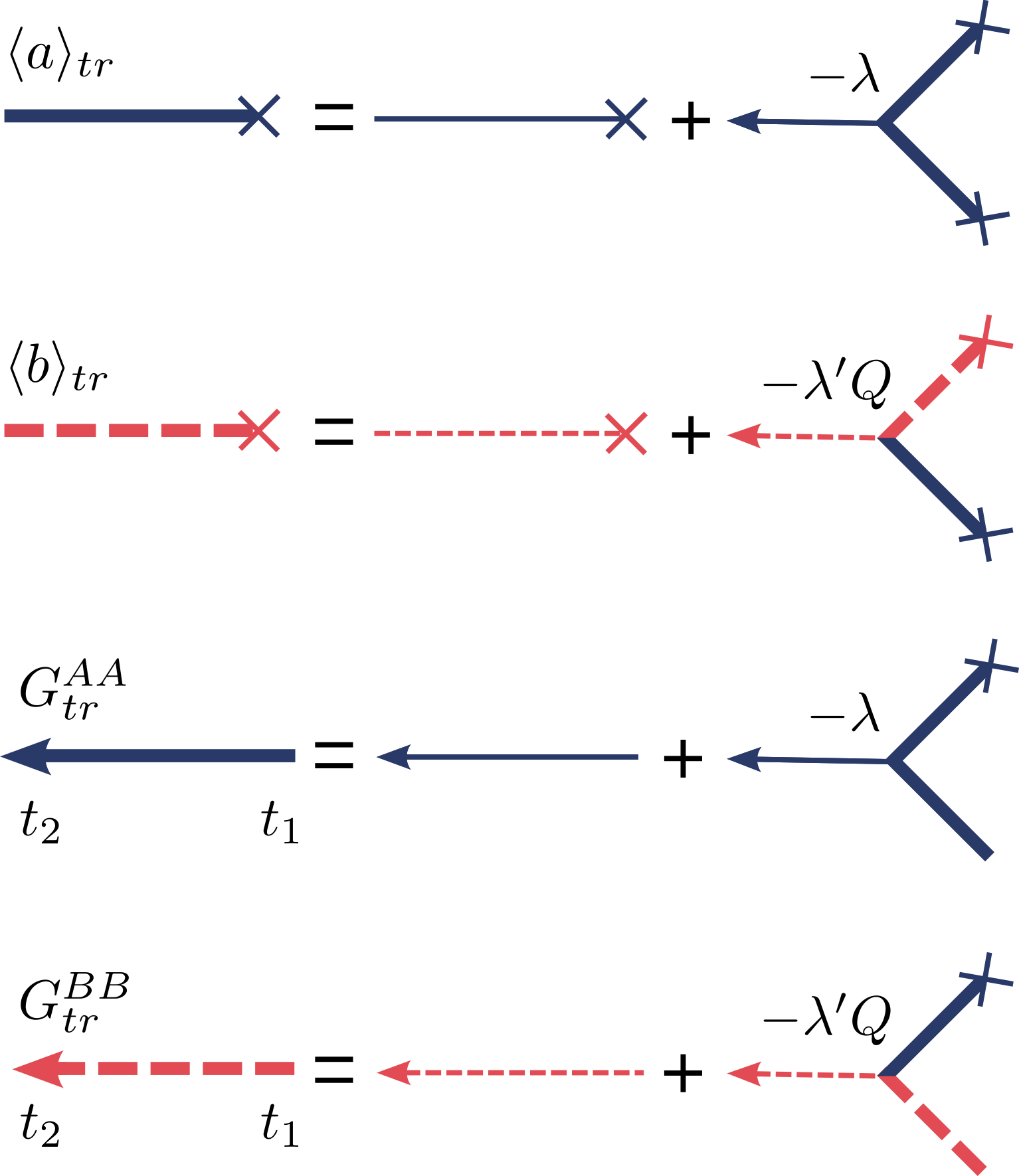}
			\caption{Diagrammatic representation of the Dyson equation for the tree-level densities $\langle a(t) \rangle_{tr}$ and $\langle b(t) \rangle_{tr}$ (upper two lines), as well as the dressed propagators $G_{tr}^{AA}$ and $G_{tr}^{BB}$ (lower two lines).}
			\label{dyson_eqs}	
		\end{center}
	\end{figure}
		
	\begin{figure}
		\begin{center}
			\includegraphics[width=90mm]{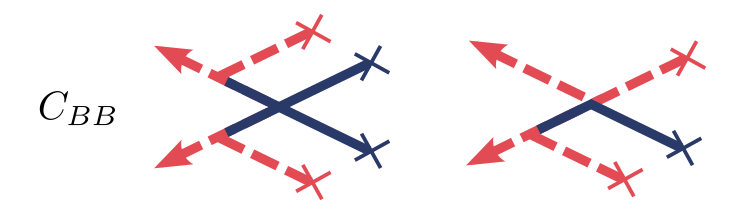}
			\caption{Tree-level diagrams for the correlation function 
				$C_{BB}$.}
			\label{tree_corr}	
		\end{center}
	\end{figure}
	
\section{Renormalization}
\label{renormalization}
\setcounter{footnote}{2}

Diagrams containing loops appear to be divergent   below $d_{c} = \sigma$ for large time limit  $t\to \infty$. These  divergencies can be handled by proper renormalization of couplings $\lambda$ and $\lambda'$. It appears that all vertices in the action (\ref{action}) renormalize identically, that leads infinite diagrammatic series for vertex renormalization to be obtained via Bethe-Salpeter equation and  to result in:
	\begin{eqnarray}
		\label{renorm_vertices_1}
		\lambda_{R}(\vec{k}=0, s) = \frac{\lambda}{1 + \lambda I_{1}(\vec{k}=0, s)} = \frac{\lambda}{1 + \lambda C_{1} \Gamma{(\epsilon/\sigma)} \, s^{-\epsilon/\sigma}},  \\
		\lambda_{R}^{\prime}(\vec{k}=0, s) = \frac{\lambda^{\prime}}{1 + \lambda^{\prime} I_{2}(\vec{k}=0, s)} = \frac{\lambda^{\prime}}{1 + \lambda^{\prime} C_{2} \Gamma{(\epsilon/\sigma)} \, s^{-\epsilon/\sigma}},
		\label{renorm_vertices_2}
	\end{eqnarray}
with $\epsilon=\sigma-d$ and
	\begin{eqnarray}
		C_{1} &=& \frac{4}{2^{d/\sigma}} \frac{\Gamma{(d/\sigma)}}{(4\pi)^{d/2}\Gamma{(d/2)} \, \sigma}, \\
		C_{2} &=& \frac{2}{(1 + \delta)^{d/\sigma}} \frac{\Gamma{(d/\sigma)}}{(4\pi)^{d/2}\Gamma{(d/2)} \, \sigma}.
	\end{eqnarray}
Quantities  $I_{1}(\vec{k}, s)$ and $I_{2}(\vec{k}, s)$ in denominators of Eqs. (\ref{renorm_vertices_1}) and (\ref{renorm_vertices_2}) are the Laplace transforms of the one-loop integrals
	\begin{eqnarray}
		I_{1}(\vec{k}, t) = 2 \int \frac{d k_{1} d k_{2}}{(2 \pi)^d} {\hat \delta}(k - k_{1} - k_{2}) e^{- k_{1}^{\sigma} t} e^{- k_{2}^{\sigma} t} , \nonumber \\
		I_{2}(\vec{k}, t)= \int \frac{d k_{1} d k_{2}}{(2 \pi)^d} {\hat \delta}(k - k_{1} - k_{2}) e^{- k_{1}^{\sigma} t} e^{- \delta k_{2}^{\sigma} t},  \nonumber
	\end{eqnarray}
where ${\hat \delta}(\dots)$ stands for  Dirac  delta-function.
	
Using standard methods we introduce normalization scale $\kappa$ and define the dimensionless coupling constants $g_{0} = \lambda \kappa^{-\epsilon}$ and  $g_{0}^{\prime} = \lambda^{\prime} \kappa^{-\epsilon}$ at  $s = \kappa^{\sigma}$, $k=0$ \cite{vernon2003}. From (\ref{renorm_vertices_1}) and (\ref{renorm_vertices_2}) we get:
	\begin{eqnarray}
		\label{gRs}
		g_{R} = \frac{g_{0}}{1 + g_{0}/g_{*}}, \qquad g_{R}^{\prime} = \frac{g_{0}^{\prime}}{1 + g_{0}^{\prime}/g_{*}^{\prime}},
	\end{eqnarray}
where the fixed points $g_{*}$ and $g_{*}^{\prime}$:
	\begin{eqnarray}
		\label{fp_1}
		g_{*} &=& \left[C_{1} \Gamma{(\epsilon/\sigma)}\right]^{-1} 
		=\frac{1}{2}(4 \pi)^{\sigma/2} \Gamma{(\sigma/2)} \epsilon + \mathcal{O}(\epsilon^{2}), \\
		g_{*}^{\prime} &=& \left[C_{2} \Gamma{(\epsilon/\sigma)}\right]^{-1} 
		=\frac{1}{2}(4 \pi)^{\sigma/2} (1 + \delta) \Gamma{(\sigma/2)} \epsilon + \mathcal{O}(\epsilon^{2}).
		\label{fp_2}
	\end{eqnarray}

Next we describe  renormalization  for the  density and correlation function of $B$ particles, introducing renormalization factors $Z_{b}$ and $Z_{b^{2}}$. In particular, the bare density relates to the renormalized one via $b_{B} = Z_{b} b_{R}$, where $Z_{b} = Z_{b}(g_{R}, g_{R}^{\prime})$ is chosen in a way that the expansions of the renormalized density have no divergences in $\epsilon$. The renormalization of correlation function is related to the square of the field associated with the $B$ particles, since the renormalization constant $Z_{b^{2}}$ is not equivalent to $(Z_{b})^{2}$ \cite{vollmayrlee2018}. In the further calculations we work with the unscaled correlation function, considering $C_{BB}(\vec{r}, t) = \langle b(\vec{r}, t) b(0, t) \rangle - \langle b(t)\rangle^{2}$, and, for convenience, in the Fourier space $\hat{C}_{BB}(\vec{k}, t) = \int d^{d} x {C}_{BB}(\vec{r}, t) e^{i \vec{k} \vec{r}}$ at zero external momenta $\vec{k} = 0$. Taking into account scaling (\ref{scal2}) we get 
	\begin{equation}\label{unscla}
		{\hat C}_{BB}(k=0,t) \sim t^{\phi-2\theta+d/\sigma}.
	\end{equation}
Therefore the bare correlation function relates to the renormalized one via $\hat{C}_{BB}^{B} = Z_{b^{2}} \hat{C}_{BB}^{R}$, where $Z_{b^{2}} = Z_{b^{2}}(g_{R}, g_{R}^{\prime})$.

Determining $Z_{b}$ and $Z_{b^{2}}$, we may find the scaling exponents  of the $B$ particle density  and  correlation function related to an anomalous dimensions \cite{vollmayrlee2018}. Our quantities of interest (density  and correlation function) must be independent of the choice of the normalization parameter $\kappa$ therefore we use dimensional analysis and obtain the RG equations:	
	\begin{flalign}
		\label{RG_eq__dens}
		\left[\sigma t \frac{\partial}{\partial t} - a_{0} d \frac{\partial}{\partial a_{0}} + \beta{(g_{R})} \frac{\partial}{\partial g_{R}} + \beta{(g_{R}^{\prime}) \frac{\partial}{\partial g_{R}^{\prime}}} + \gamma_{b}(g_{R}, g_{R}^{\prime})\right] b_{R}\left(t, a_{0}, g_{R}, g_{R}^{\prime}; \kappa\right) = 0,
	\end{flalign}
	
	\begin{flalign}
		\label{RG_eq__corr}
		\left[\sigma t \frac{\partial}{\partial t} - a_{0} d \frac{\partial}{\partial a_{0}} + \beta{(g_{R})} \frac{\partial}{\partial g_{R}} + \beta{(g_{R}^{\prime}) \frac{\partial}{\partial g_{R}^{\prime}}} + \gamma_{b^{2}}(g_{R}, g_{R}^{\prime}) - d\right] \hat{C}_{BB}^{R}\left(t, a_{0}, g_{R}, g_{R}^{\prime}; \kappa\right) = 0,
	\end{flalign}
with $\beta$-functions  defined as, 
	\begin{eqnarray}
		\label{beta_1}
		\beta{(g_{R})} &\equiv& \kappa \frac{\partial g_{R}}{\partial \kappa} = - \epsilon g_{R} + \frac{\epsilon}{g_{*}} g_{R}^{2}, \\
		\beta{(g_{R}^{\prime})} &\equiv& \kappa \frac{\partial g_{R}^{\prime}}{\partial \kappa} = - \epsilon g_{R}^{\prime} + \frac{\epsilon}{g_{*}^{\prime}} {g_{R}^{\prime}}^{2}.
		\label{beta_2}
	\end{eqnarray}
Fixed points of $\beta$-functions above are given by (\ref{fp_1}) and (\ref{fp_2}) respectively, they are stable for $d < d_c = \sigma$. $\gamma$-functions in (\ref{RG_eq__dens}) and (\ref{RG_eq__corr}) define the anomalous dimensions when calculated the fixed points $g_{*}$ and $g_{*}^{\prime}$:
	\begin{eqnarray}\label{gzb}
		\gamma_{b}(g_{R}, g_{R}^{\prime}) = \kappa \frac{\partial}{\partial \kappa} \ln{Z_{b}},
	\end{eqnarray}
	\begin{eqnarray}\label{gzb2}
		\gamma_{b^{2}}(g_{R}, g_{R}^{\prime}) = \kappa \frac{\partial}{\partial \kappa} \ln{Z_{b^{2}}}.
	\end{eqnarray}

Solving Eqs. (\ref{RG_eq__dens}) and  (\ref{RG_eq__corr}) by the method of characteristics, we find the following asymptotic solutions:
	\begin{eqnarray}
		\label{dens_asymptotic}
		b_{R}\left(t, a_{0}, g_{R}, g_{R}^{\prime}; \kappa\right) \sim (\kappa^{\sigma} t)^{-\gamma_{b}^{*}/\sigma} b_{R}\left(\kappa^{-\sigma}, a_{0} (\kappa^{\sigma} t)^{d/\sigma}, \tilde{g}_{R}, \tilde{g}_{R}^{\prime}; \kappa\right),
	\end{eqnarray}
	\begin{eqnarray}
		\label{corr_asymptotic}
		\hat{C}_{BB}^{R}\left(t, a_{0}, g_{R}, g_{R}^{\prime}; \kappa\right) \sim (\kappa^{\sigma} t)^{d/\sigma-\gamma_{b^{2}}^{*}/\sigma} \hat{C}_{BB}^{R}\left(\kappa^{-\sigma}, a_{0} (\kappa^{\sigma} t)^{d/\sigma}, \tilde{g}_{R}, \tilde{g}_{R}^{\prime}; \kappa\right),
	\end{eqnarray}
where $\gamma_{b}^{*} = \gamma_{b}(g_{*}, g_{*}^{\prime})$, $\gamma_{b^{2}}^{*} = \gamma_{b^{2}}(g_{*}, g_{*}^{\prime})$, $\tilde{g}_{R}$ and $\tilde{g}_{R}^{\prime}$ are the running couplings, which go at $t \to \infty$ to the fixed values $g_{*}$ and $g_{*}^{\prime}$.	As follows, the asymptotic time dependence of the bare density and  bare correlation function is determined by the renormalized $a_{0}$ and  by the anomalous dimensions $\gamma_{b}^{*}$ and $\gamma_{b^2}^{*}$ correspondingly (similarly as in \cite{vollmayrlee2018}).

Next, we consider the results for the $B$ particle density and correlation function, taking into account the one-loop contributions.

\section{Results of renormalization}
\label{results}
\setcounter{footnote}{3}

\subsection{The $B$ particle density}
\label{density}
\setcounter{footnote}{4}

Results for one-loop diagram contributing to density of $B$ particles are presented in Appendix~\ref{ap1}. Expanding them in powers of $\epsilon$, then  collecting contributions divergent at $\epsilon\to 0$ and combining with tree-level result we get 
	\begin{eqnarray}
		\label{bare_density}
		b_{B} = \frac{b_{0}}{(a_{0} \lambda t)^{Q z} }\left[1 + \lambda t^{\epsilon/\sigma} \left(\frac{{\cal A}(z)}{\epsilon^{2}} + \frac{{\cal B}(z)}{\epsilon} + \ldots \right)\right].
	\end{eqnarray}
where  we have introduced  $z \equiv \lambda^{\prime} / \lambda = g_{0}^{\prime} / g_{0}$. ${\cal A}(z)$ and ${\cal B}(z)$ in (\ref{bare_density}) read:
	\begin{eqnarray}{\cal A}(z) &=& 2 \sigma Q z \frac{1}{(4 \pi)^{d/2}} \frac{\Gamma(d/\sigma)}{\Gamma(d / 2)} \left(\frac{z}{(1 + \delta)^{d/\sigma}} - \frac{2}{2^{d/\sigma}}\right), \nonumber \\
		{\cal B}(z) &=&  Q z \frac{1}{(4 \pi)^{d/2}} \frac{\Gamma(d/\sigma)}{\Gamma(d / 2)} \left(\frac{5}{2^{d/\sigma - 1}} - \frac{2 z}{(1 + \delta)^{d/\sigma}} + \frac{Q z}{(1 + \delta)^{d/\sigma - 1}} f(\delta)  \right)
	\end{eqnarray}
with
	\begin{eqnarray}
		\label{f(delta)}
		f\left(\delta\right) = 1 + 2 \delta \left[\ln{\left(\frac{2}{\delta + 1}\right)- 1}\right] + \left(1 - \delta^{2}\right)\left[\mathrm{Li}_{2}\left(\frac{\delta - 1}{\delta + 1}\right) - \frac{\pi^{2}}{6}\right],  
	\end{eqnarray}
where $\mathrm{Li}_{2}(...)$ is the dilogarithm function. Function $f(\delta)$ has   the same form is in the case with ordinary diffusion  \cite{rajesh2004, vollmayrlee2018}. The contribution  proportional to $1/\epsilon^{2}$ in (\ref{bare_density}) can be neglected  since we are interested by behaviour at the fixed point, where:
	\begin{eqnarray}
		\label{z*}
		z* = \frac{g_{*}^{\prime}}{g_{*}} = 2 \left(\frac{1 + \delta}{2}\right)^{d/\sigma} 
		\underset{d \to \sigma}{=\mathrel{\mkern-2.5mu}=}
		1 + \delta + \mathcal{O}\left(\epsilon\right).
	\end{eqnarray} 
Therefore   ${\cal A}(z)$  vanishes at $z \to z^{*}$, since it is proportional to $\propto (z - z^{*})$. 
	
Considering the contribution under renormalization proportional to $1/\epsilon$, we expand ${\cal B}(z)$ in powers of $z - z^{*}$, leave only the leading terms we get:

	\begin{flalign}
		\label{Bz}
		{\cal B}(z^{*}) = \frac{1}{(4 \pi)^{\sigma/2} \Gamma{(\sigma/2)}} \left[3 Q (1 + \delta) + Q^{2} (1 + \delta)^{2} f\left(\delta\right)\right] + \mathcal{O}\left(\epsilon, z - z^{*}\right).
	\end{flalign}
	
Next, we apply standard steps  substituting $t \to \kappa^{-\sigma}$, $a_{0} \to  a_{0} (\kappa^{\sigma} t)^{d/\sigma}$, $\lambda \to \kappa^{\epsilon} g_{0} = \kappa^{\epsilon} (g_{R} + g_{R}^{2}/g_{*} + \ldots)$ in Eq. (\ref{bare_density}) and expanding obtained expression in powers of the renormalized couplings $g_{R}$ and $g_{R}^{\prime}$. We find  the following expression for the bare density in the linear order for the renormalized constants:
	\begin{eqnarray}
		\label{dens__gR_exp}
		b_{B} = \frac{b_{0}}{(a_{0} g_{R} t^{d/\sigma})^{Q g_{R}^{\prime}/g_{R}}} \left[1 + \frac{{\cal B}(z^{*})}{\epsilon} g_{R} - \frac{Q}{g_{*}} g_{R}^{\prime} + \ldots\right].
	\end{eqnarray}
As can be seen from the expansion (\ref{dens__gR_exp}), it is necessary to renormalize the field due to the appearance of the term proportional to $1/\epsilon$, which allows to identify the renormalization constant $Z_{b}$. Therefore in the linear order to the renormalized couplings $g_{R}$ and $g_{R}^{\prime}$ and $1/\epsilon$  constant $Z_{b}$ is written as:
	\begin{eqnarray}
		\label{zb}
		Z_{b} = 1 + \frac{{\cal B}(z^{*})}{\epsilon} g_{R} - \frac{Q}{g_{*}} g_{R}^{\prime} + \ldots,
	\end{eqnarray}
and consequently from (\ref{zb}) we obtain 
	\begin{eqnarray}
		\gamma_{b} = - {\cal B}(z^{*}) g_{R} + \frac{2 Q}{(4\pi)^{\sigma/2} \Gamma(\sigma/2)} g_{R}^{\prime}.
	\end{eqnarray}
Calculated at the fixed point values $g_{*}$ and $g_{*}^{\prime}$,  (\ref{fp_1}) and (\ref{fp_2}), it gives us the following result
	\begin{eqnarray}
		\label{gamma_b}
		\gamma_{b}^{*} = - \left[\frac{1}{2} \left(\frac{1 + \delta}{2 - p}\right) + \frac{1}{2} \left(\frac{1 + \delta}{2 - p}\right)^{2} f\left(\delta\right)\right]\epsilon + \mathcal{O}\left(\epsilon^{2}\right),
	\end{eqnarray}
where we have used $Q = 1/(2-p)$. Comparing the obtained anomalous dimension (\ref{gamma_b}) for the two-species reaction-diffusion system with L\'evy flights to the result for the short-range diffusion hops, we find exactly the same exponent resulting from the renormalization of the field associated with the $B$ particles up to value of $\epsilon$ \cite{vollmayrlee2018}. 
	
And finally, the renormalized density has the following behavior:
	\begin{eqnarray}
		\label{renorm_dens}
		b_{R} = Z_{b}^{-1} b_{B} \sim t^{-d Q z^{*} / \sigma - \gamma_{b}^{*} / \sigma} = t^{- \theta},
	\end{eqnarray}
where the first term $d Q z^{*} / \sigma$ with fixed point value (\ref{z*}) corresponds to the Smoluchowski exponent  (for description of this approximation for our model see Appendix \ref{ap0}, for Smoluchowski approximation prediction for $\theta$ see (\ref{smoluch_exp})), while the second term is a result of the renormalization of the field associated with the $B$ particles. 	Therefore we have the expression for $\theta$ in the first order in $\epsilon=\sigma-d$: 
	\begin{flalign}
		\label{theta}
		\theta =  \frac{{2} \, d}{\sigma (2-p)} \left(\frac{1 + \delta}{2}\right)^{d/\sigma} +\frac{1}{2\sigma}\left[ \left(\frac{1 + \delta}{2 - p}\right) +  \left(\frac{1 + \delta}{2 - p}\right)^{2} f\left(\delta\right)\right](\sigma-d) .
	\end{flalign}
Note that  $\sigma$ in the denominator in (\ref{theta}) leads to increasing $\theta$ with decreasing of $\sigma$, therefore to slower decay of density $\langle b\rangle$ for L\'evy flight distribution with larger probability for long hops, see Fig.~\ref{Levydens}. Moreover, comparing  (\ref{theta})  with result for ordinary diffusion \cite{vollmayrlee2018} we can see that our result (\ref{theta})  can be obtained from \cite{vollmayrlee2018} with  substitution $d_{\rm eff}=2d/\sigma$ instead $d$. Since we consider $\sigma<2$ it leads to the conclusion that in system where both traps and target particles perform L\'evy flight, probability for $B$ particles to survive is larger.

	\begin{figure}[t!]
		\begin{center}
			\includegraphics[width=90mm]{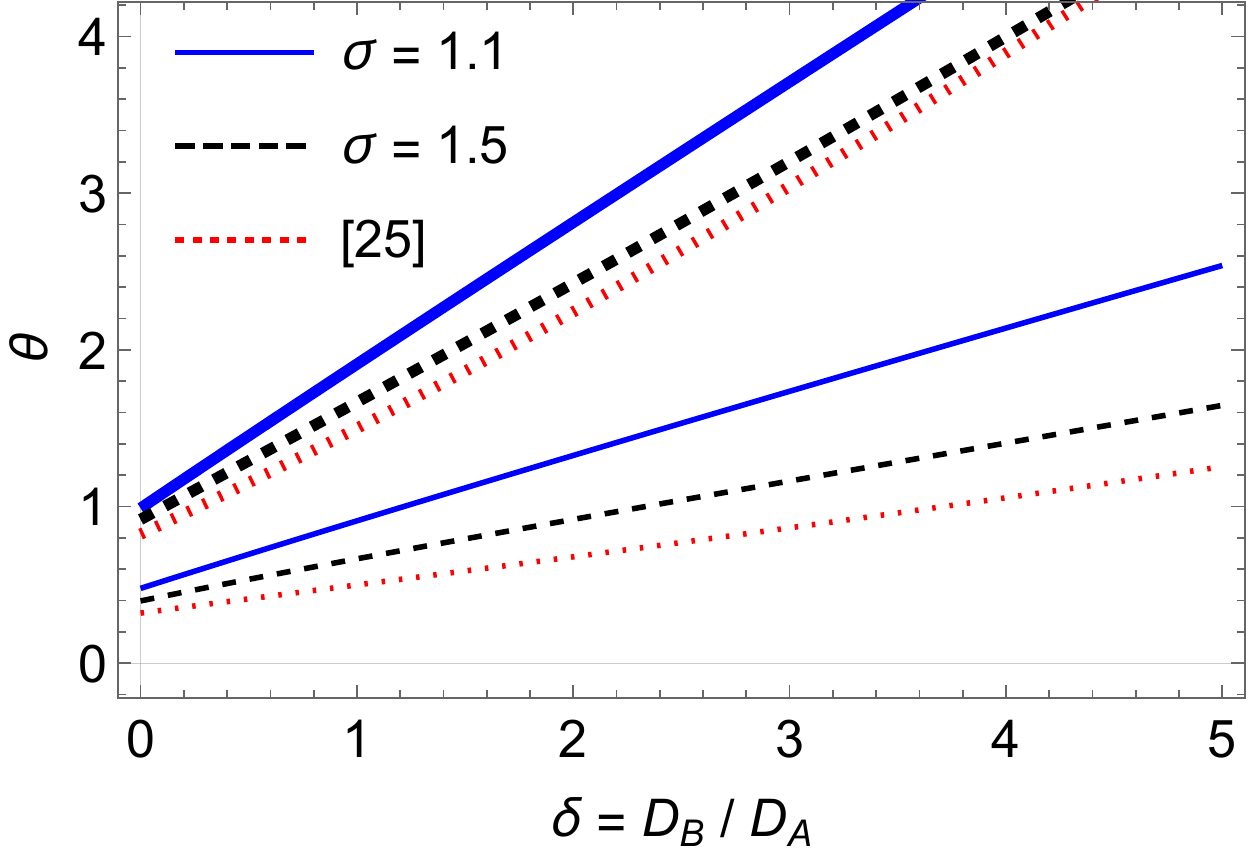}
			\caption{The $B$ particles density decay exponent $\theta$ as a function of the diffusion constants ratio $\delta = D_{B} / D_{A}$ in the one-dimensional case $d = 1$. The upper set of curves (thick) correspond to  $A + A \to A$ (coalescence), while the lower ones (thin) correspond to $A + A \to 0$ (annihilation), respectively. Solid curves are plotted for $\sigma = 1.1$ and the  dashed ones are plotted for $\sigma = 1.5$, while the  dotted curves is a result for the case of the diffusion hops \cite{vollmayrlee2018}.}
			\label{Levydens}	
		\end{center}
	\end{figure}

	\begin{figure}[t!]
		\begin{center}
			\includegraphics[width=90mm]{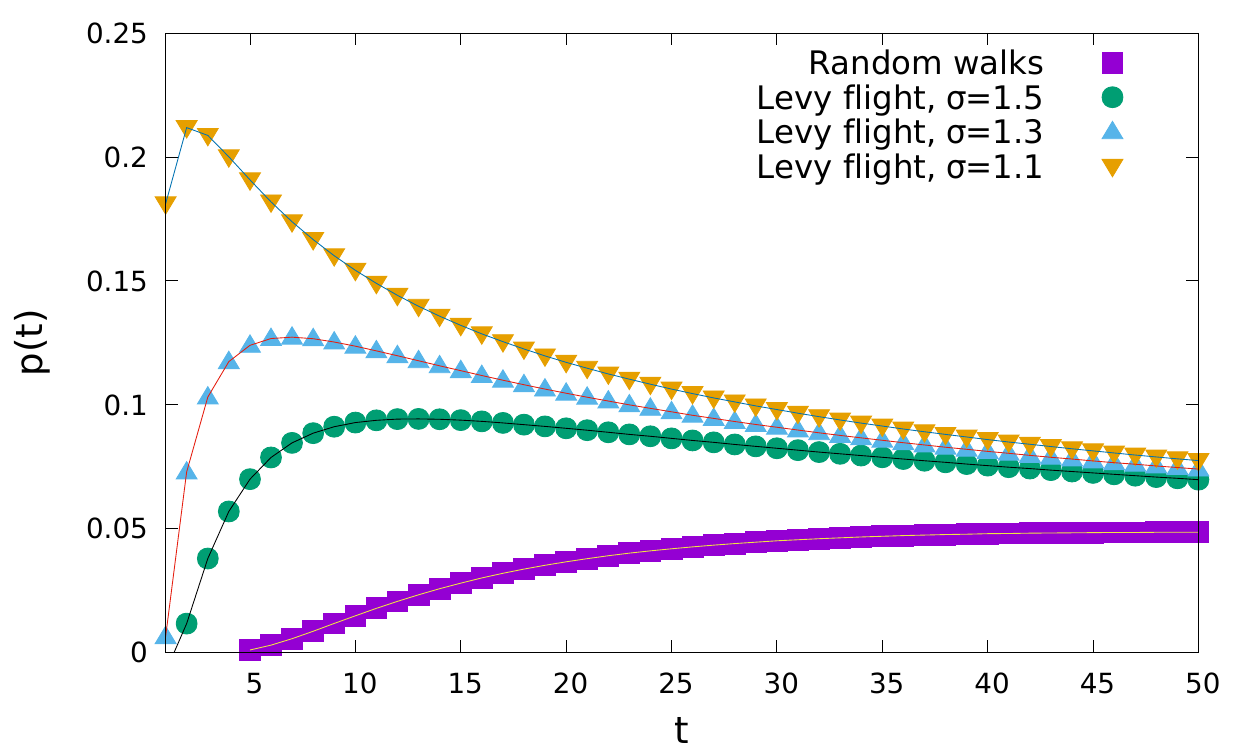}
			\caption{An estimate for the probability for two particles, initially separated by distance $l_0=10$ on the one-dimensional lattice, to  to meet at time $t$. 
				The cases, when two particles are performing simple random walks and L\'evy flights with exponents  $\sigma=1.1$, $\sigma=1.3$  and $\sigma=1.5$ are shown.	}
			\label{probLevy}	
		\end{center}
	\end{figure}
	
Higher survival probability for $B$ particles in the case of L\'evy flights in comparison with that for ordinary transitions can be comprehended as a result of a direct consequence of optimization of encounter rate in L\'evy flight process. Indeed, let us evaluate  the probability for two randomly walking particles A and B to meet after performing $t$ steps, if initially they are separated by distance $L_0=l_0$ on an one-dimensional lattice.	At a time $t$, each particle performs a jump to the left or to the right with probability $1/2$. This can result either in increasing the distance between them ($L_t=L_{t-1}+2$ if two particles jump in opposite direction),	in decreasing distance between them ($L_t=L_{t-1}-2$ if both are jumping towards each other), or keeping the same distance ($L_t=L_{t-1}$ if both are going to the left or right simultaneously). The probabilities of these three cases are thus correspondingly $p_{+}=p_{-}=1/4$, $p_0=1/2$. Let us denote by $t_{+}$, $t_{-}$ and $t_0$ the number of times, when particles  perform the mutual jumps of each those three types, so that $t=t_{+}+t_{-}+t_0$. The particles meet, if distance between them $L_t=0$, thus resulting in condition: $l_0+2t_{+}-2t_{-}=0$.The probability, that particles will meet after $t$ steps thus is given by
	\begin{equation}
		P(t,t_{+},t_{-})=\frac{t!}{t_{+}!t_{-}!(t-t_{+}-t_{-})!}\left(\frac{1}{4}\right)^{t_{+}}\left(\frac{1}{4}\right)^{t_{-}}\left(\frac{1}{2}\right)^{t-t_{+}-t_{-}} {\rm with} \phantom{5} t_{-}=\frac{l_0}{2}+t_{+}. \label{pprw}
	\end{equation}
Finally,  performing the sum over $t_{+}$ in the   last equation, we  obtain the probability $p(t)$ for two random walkers, presented in Fig. \ref{probLevy} for the case $l_0=10$. Let us generalize this expression for the case, when both particles are performing L\'evy flight. The averaged length $l_{{\rm av}}$ of each jump can be estimated on the basis of truncated L\'evy flights	\cite{Mantegna94,Ghaemi09}. Thus, following the ideas developed above for simplified random walks, the increase of the distance between them in the case when both particles jump in opposite direction is  ($L_t=L_{t-1}+2l_{{\rm av}}$, and correspondingly the decrease of distance for particles jumping towards each other is $L_t=L_{t-1}-2l_{{\rm av}}$.  As a result, this leads to substituting $l_0$ by $l_0/l_{{\rm av}}$ in expression (\ref{pprw}). For example, using $l_{{\rm max}}=15000$, we obtain from distribution of truncated L\'evy flights: $l_{{\rm av}}(\sigma=1.5)=1.94$, $l_{{\rm av}}(\sigma=1.3)=2.61$,  $l_{{\rm av}}(\sigma=1.1)=4.33$. The corresponding estimates are presented in Fig.~\ref{probLevy}. The position of maximum of $P(t)$ is shifted towards the smaller $t$ values with decreasing the parameter $\sigma$, which clearly supports the intuitive understanding  of increasing the encountering rate for particles performing L\'evy flight as compared with ordinary random walks.

\subsection{The $B$ particle correlation function}
\label{correlation_function}
\setcounter{footnote}{5}
	
The tree-level diagrams (see Fig.~\ref{tree_corr}) together with the one-loop diagrams (see the Appendix~\ref{ap2}) lead to the following result for the  bare $B$ particle correlation function in the large $a_{0}$ limit:
	\begin{eqnarray}
		\label{bare_corr_func}
		\hat{C}_{BB}^{B} = \frac{b_{0}^{2} h(Q) t}{(a_{0} \lambda t)^{2 Q z}} 
		z\lambda^{\prime}\left[1 + \lambda t^{\epsilon / \sigma} \left(\frac{2{\cal A}(z)}{\epsilon^{2}} + \frac{{\cal C}(z) }{\epsilon} + \ldots\right)\right],
	\end{eqnarray}
where $h(Q) = Q (1 - 2 Q / 3)$. We can omit  contribution proportional to $1 / \epsilon^{2}$  with the same reason as for the density, while collecting the terms proportional to $1 / \epsilon$ from (\ref{F1}) -- (\ref{F4}) in the Appendix~\ref{ap2} we get
	\begin{eqnarray}
		\label{Dz}
		{\cal C}(z^{*}) = \frac{2}{(4 \pi)^{\sigma/2} \Gamma{(\sigma/2)}} \left[- \frac{9 - 13 Q}{3(3 - 2 Q)} \right] +2{\cal B}(z^*)+ \mathcal{O}\left(\epsilon, z - z^{*}\right).
	\end{eqnarray}
Then we repeat steps we performed for the density in order to find the renormalization constant $Z_{b^{2}}$ with result 
	\begin{eqnarray}
		Z_{b^{2}} = 1 + \left(\frac{{\cal C}(z^{*}) }{\epsilon} - \frac{1}{g_{*}}\right)g_{R} + 2\left(\frac{1}{g_{*}^{\prime}} - \frac{Q}{g_{*}}\right)g_{R}^{\prime}.
	\end{eqnarray}
Therefore from (\ref{gzb2}) we have:
	\begin{flalign}
		\gamma_{b^{2}} = \left(-{\cal C}(z^{*})  + \frac{2}{(4 \pi)^{\sigma/2} \Gamma(\sigma/2)} \right)g_{R} + \frac{4}{(4 \pi)^{\sigma/2} \Gamma(\sigma/2)} \left(Q - \frac{1}{1 + \delta} \right)g_{R}^{\prime}.
	\end{flalign}
Finally, $\gamma_{b^{2}}$ at the fixed points $g_{*}$ and $g_{*}^{\prime}$ defined in eqs. (\ref{fp_1}) and (\ref{fp_2}) and $Q = 1/(2-p)$ gives us the following result
	\begin{eqnarray}
		\label{gamma_b2}
		\gamma_{b^{2}}^{*} = - \left[\frac{7}{12 - 9 p} + \frac{1 + \delta}{2 - p} + \left(\frac{1 + \delta}{2 - p}\right)^{2} f\left(\delta\right) \right]\epsilon + \mathcal{O}\left(\epsilon^{2}\right).
	\end{eqnarray}
Therefore the renormalized $B$ particle correlation function has the following leading behavior with time:
	\begin{eqnarray}
		\label{renorm_corr_func}
		\hat{C}_{BB}^{R} (k = 0) \sim t^{d/\sigma - 2 d Q z^{*} / \sigma - \gamma_{b^{2}}^{*} / \sigma}.
	\end{eqnarray}
Comparing obtained behaviour (\ref{renorm_corr_func}) with scaling (\ref{unscla}) we find that the exponent $\phi$ reads as:
	\begin{eqnarray}
		\label{phi}
		\phi = \frac{2}{\sigma}\left(\gamma_{b}^{*} - \frac{1}{2} \gamma_{b^{2}}^{*}\right) =	
		\frac{7}{\sigma(12 - 9 p)} \epsilon + \mathcal{O}\left(\epsilon^{2}\right).
	\end{eqnarray}
	
Inverse dependence of the result (\ref{phi}) on $\sigma$  (see also Fig.~\ref{Levycorr}) shows that in system with L\'evy flights characterizing  by smaller value of $\sigma$, $B$ particles are more correlated in time, that in system with larger parameter of L\'evy flights. As a consequence we see that in a system with ordinary diffusion target particles in time are less correlated than in a system with L\'evy flights. Moreover similarly as in the case of exponent for density decay  comparing  (\ref{phi})  with result for ordinary diffusion \cite{vollmayrlee2018} we can see that it  can be obtained  from $\phi$ for ordinary diffusion  substituting $d$ by $d_{\rm eff}=2d/\sigma$.
	
	\begin{figure}[t!]
		\begin{center}
			\includegraphics[width=90mm]{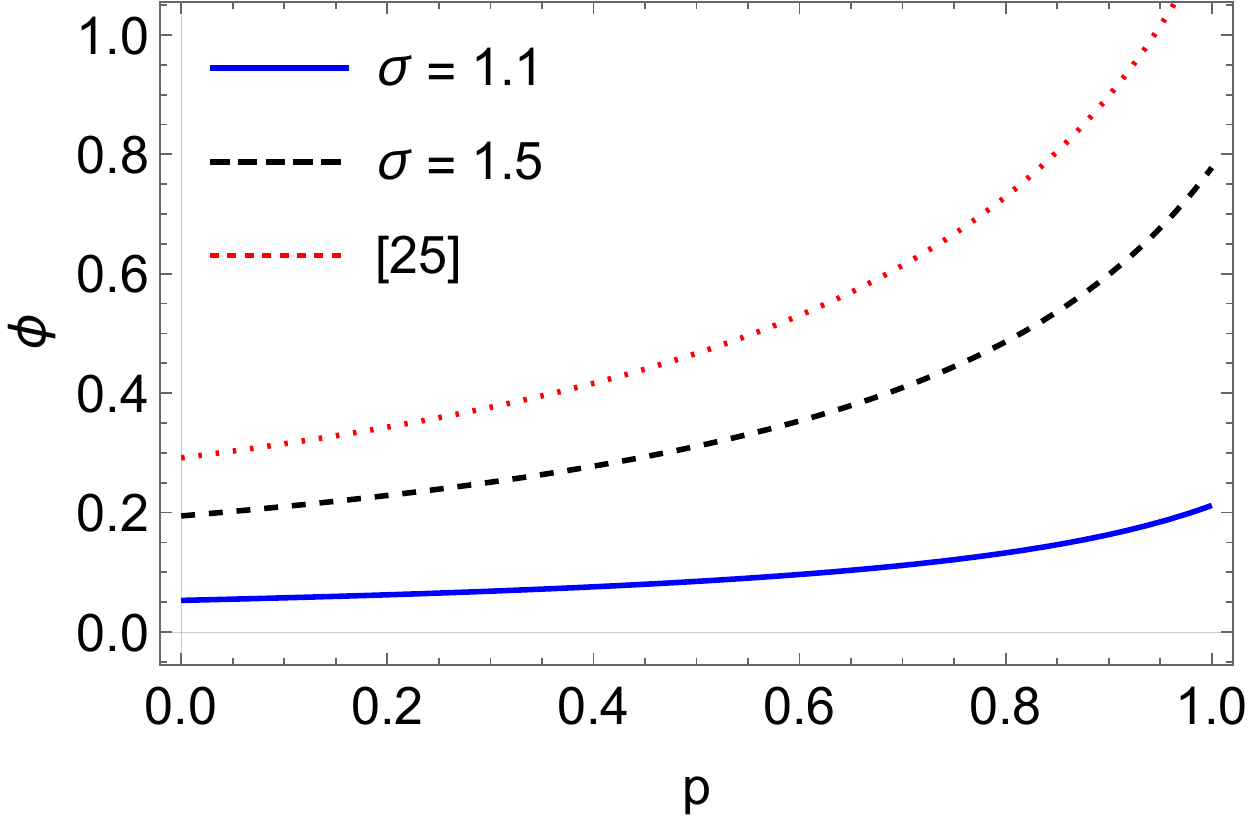}
			\caption{The $B$ particles correlation function exponent $\phi$ as a function of $p$ in the one-dimensional case $d = 1$. Solid curve is plotted for $\sigma = 1.1$ and the  dashed one is plotted for $\sigma = 1.5$, while the dotted curve is a result for the case of the diffusion hops \cite{vollmayrlee2018}.}
			\label{Levycorr}	
		\end{center}
	\end{figure}

\setcounter{footnote}{6}
	
\section{Numerical results for the one-dimensional case}
\label{V}
	\begin{figure}[t!]
		\begin{center}
			\includegraphics[width=90mm]{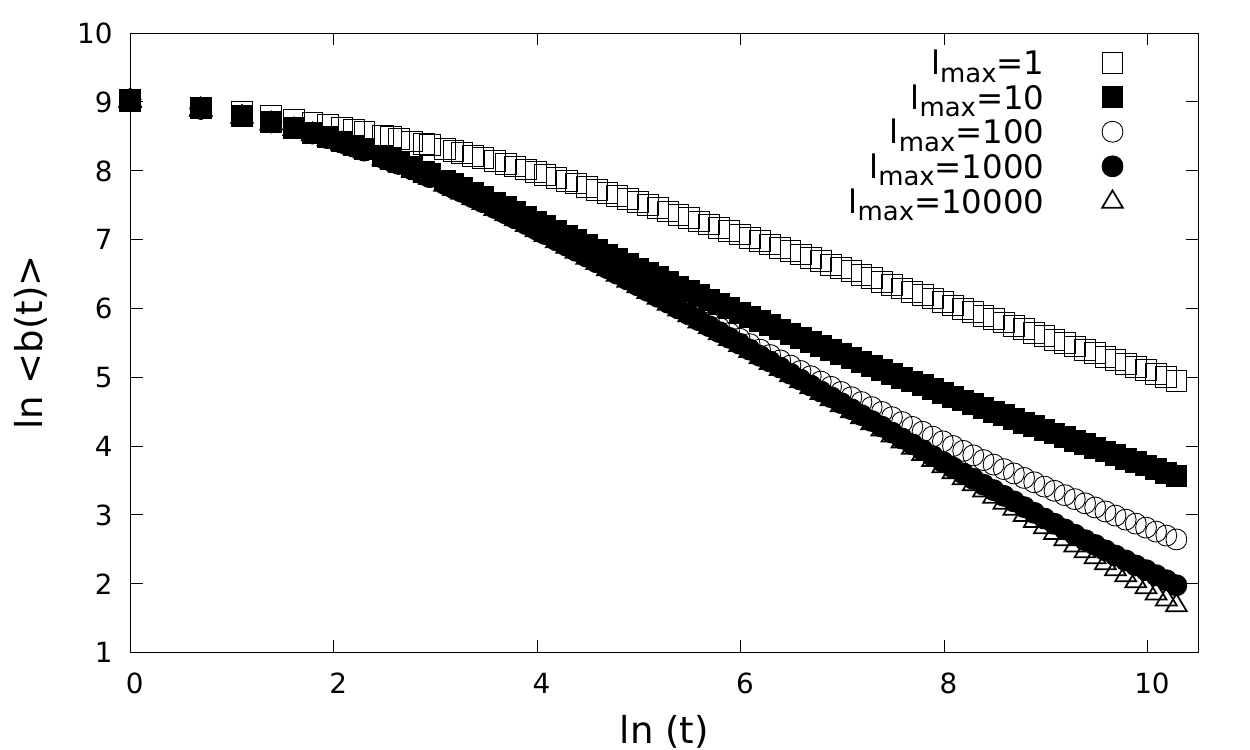}
			\caption{Averaged concentration of $B$ particles as a function of $t$ in the double logarithmic scale at fixed parameters $\delta=1$, $\sigma=1.1$ and coalescence probability $p=0$ at various $l_{{\rm max}}$.
				A crossover from an asymptotic regime of ordinary diffusion (at $l_{{\rm max}}=1$)  towards L\'evy statistics is observed.
			}
			\label{Levy21kmax}	
		\end{center}
	\end{figure}
In this section we complete our analytical finding of density decay exponent for $B$ particles by  estimate obtained from numerical simulations.
	
To analyze the peculiarities of the two-species reaction-diffusion system within the frames of computer simulations, we start with discrete representation of a model   based on an  one-dimensional lattice (chain) containing $N=10^5$ sites. At each discrete moment of time  time $t$, the labels $n_A(i,t)$ and $n_B(i,t)$ are prescribed to $i$th site, which equal $n$  when the site contains $n$ particles of the type $A$ or $B$ respectively, or $0$ otherwise. Densities of $ A$ and $B$ particles  are thus given by:
	\begin{equation}
		a(t)=\frac{1}{N}\sum_i n_A(i,t), \,\,b(t)=\frac{1}{N}\sum_i n_B(i,t).
	\end{equation}
At the moment $t=0$, we set $a(0)=b(0)=0.1$ as initial densities of particles $A$ and $B$, settled on randomly chosen sites.  
	
We apply the synchronous version of cellular automaton  updating algorithm, where one time step implies a sweep throughout the whole system. The time update $t-1 \to t$ consists of two steps. At a first step,  we check the state of each $i$th site. If it contains a particle ($n_C(i,t)>0$ with $C=\{A,B\}$),  the particle makes a jump of the length $l$ to the right or to the left with $l$ obeying a L\'evy statistics. As a result of such jump, one has $n_C(i,t)=n_C(i,t-1)-1$ and  $n_C(i\pm l,t)=n_C(i\pm l,t-1)+1$. Note that since the system under consideration is finite, a cut-off $l_{{\rm max}}$ of the maximal length of a jump is introduced, so that the lengths are taken from distribution function in a form 
	\begin{equation}
		p(l)=\frac    { l^{-(d+\sigma)}  } {  \sum_{l=1}^{l_{{\rm max}} }  l^{-(d+\sigma)}  }, \label{disLevy}
	\end{equation}   
which corresponds to the so-called truncated L\'evy flights \cite{Mantegna94,Ghaemi09}. 
By tuning the parameter $l_{{\rm max}}$, we observe a crossover from an asymptotic regime of ordinary diffusion (at $l_{{\rm max}}\ll T$)  towards L\'evy statistics. Note also, that when the $A$ and $B$ diffusion constants are equal ($\delta=1$), the particles of both types jump simultaneously at each moment of time $t$. Otherwise, the particles with smaller diffusion constants skip some time moments. For example, in the case $\delta=2$ diffusion coefficient of $B$ particles is two times larger and they jump two times faster comparing to $A$ particles. It can be realized in a way, that at even values of $t$ both $A$ and $B$ are making jumps, whereas at odd values of $t$ only $B$ are jumping and $A$ are staying at the same positions they occupied at time $t-1$. 
	
At the second step of time update,  the reaction rules are applied:
\begin{itemize}
	\item  if $n_A(i,t)=1$ and $n_B(i,t)>1$, then  $n_B(i,t)=n_B(i,t)-1$;
	\item  if $n_A(i,t)=2$,  then  $n_A(i,t)=1$  with probability $p$ (coalescence) or $n_A(i,t)=0$ with probability $1-p$ (annihilation).  
\end{itemize}
The ensemble averaging $\langle (\ldots) \rangle$  are performed over 1000 replicas.

\begin{figure}[t!]
	\begin{center}
		\includegraphics[width=90mm]{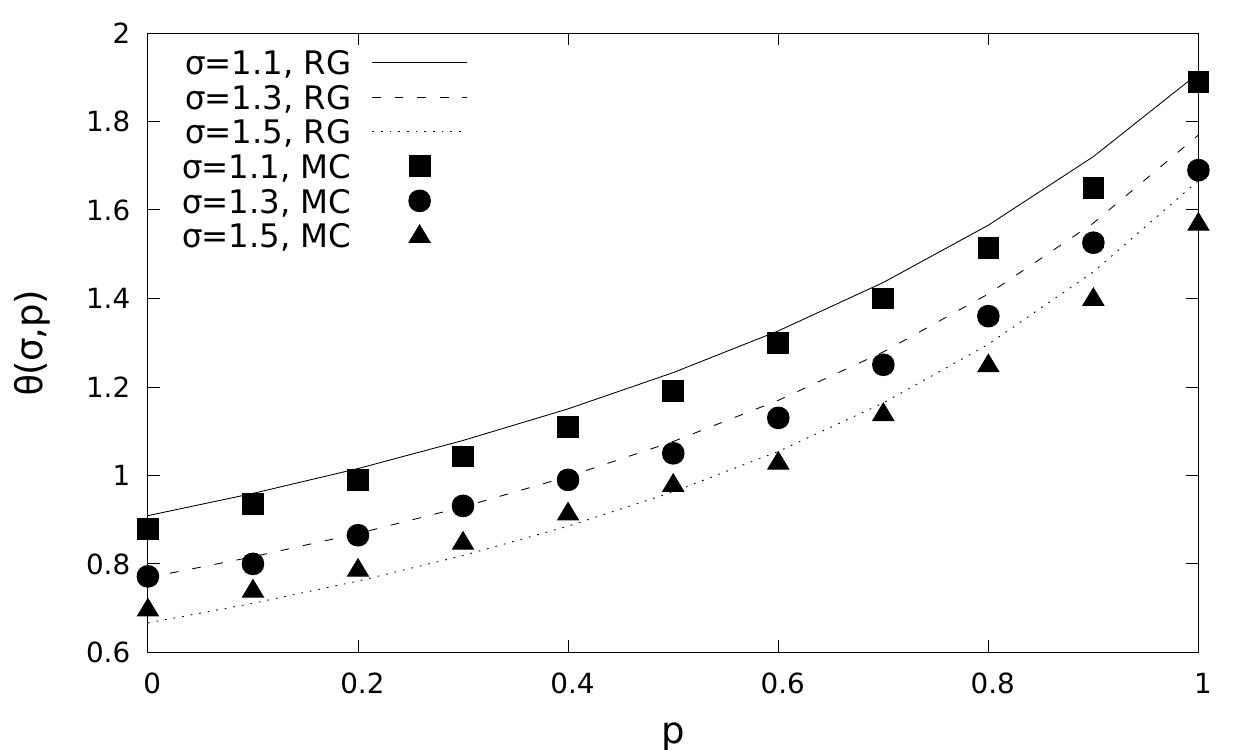}
		\caption{Critical exponent $\theta$  as function of  coalescence probability $p$ at $\delta=1$ and various $\sigma$. Lines: RG results, symbols: results of numerical simulations}
		\label{thetaeq}	
	\end{center}
\end{figure}

\begin{figure}[t!]
	\begin{center}
		\includegraphics[width=90mm]{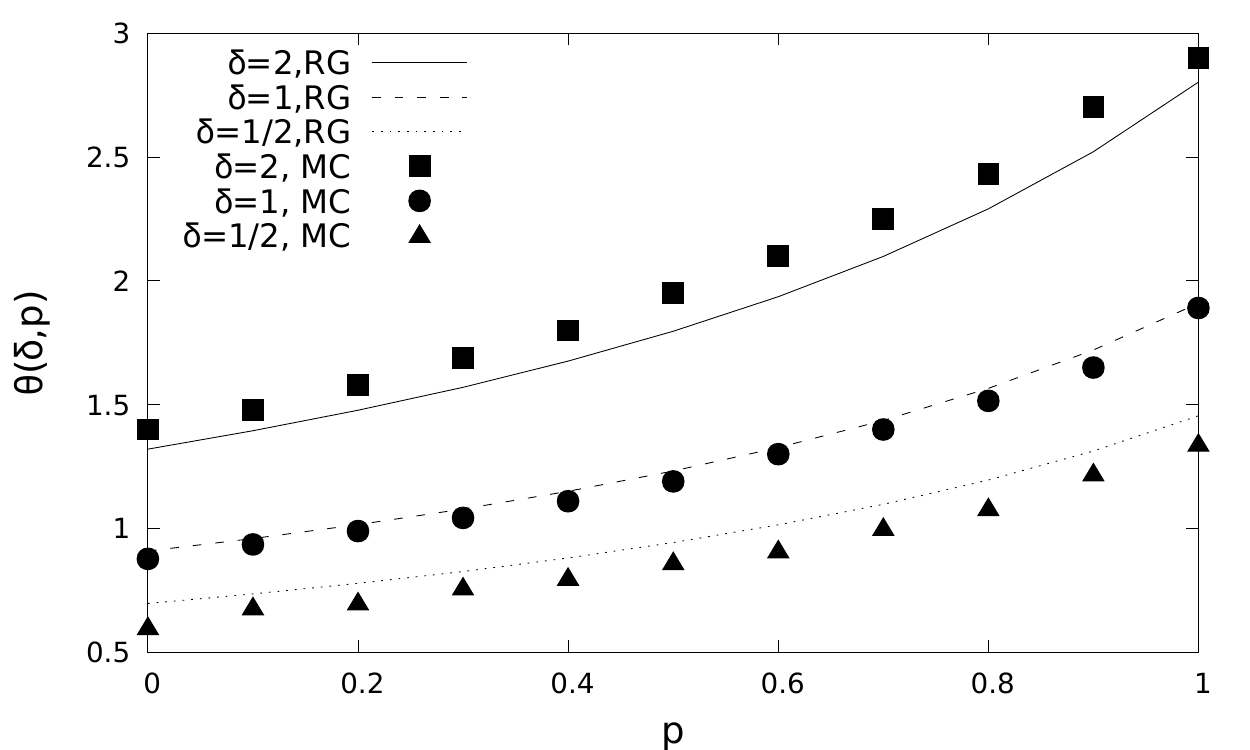}
		\caption{Critical exponent $\theta$  as a function of  coalescence probability $p$ at $\sigma=1.1$ and various $\delta$. Lines: RG results, symbols: results of numerical simulations}
		\label{thetadelta}	
	\end{center}
\end{figure}

On Fig. \ref{Levy21kmax} we present results for the time dynamics of $\langle b(t) \rangle $ for $\sigma=1.1$ at various parameter $l_{{\rm max}}$. The estimates for critical exponent $\theta$ are obtained by linear least-square fitting to the form (\ref{scal1}) with varying lower cutoff for the number of time steps; the sum of squares of normalized deviation from the regression line divided by the number of degrees of freedom served as a test of the goodness of fit. At $l_{{\rm max}}=1$, we restore the known result for $\theta(p=0)=0.5$ of ordinary diffusion. As one can immediately observe from the  Fig. \ref{Levy21kmax}, increasing of the maximum jump length causes the density of survived particles to decrease faster.   To obtain the estimates for the exponents $\theta$ in the case of L\'evy statistics, we exploited the values $l_{{\rm max}}=15000$. Our data for $\theta$ as function of coalescence probability $p$ at various parameters $\sigma$ and $\delta$ are presented correspondingly on Figs. \ref{thetaeq} and \ref{thetadelta} in comparison of corresponding analytical results. At each fixed value of coalescence probability $p$, the exponents increase with increasing the parameter $\sigma$. Let us recall, that the smaller is $\sigma$, the larger is the probability for very long jumps to occur and thus, this qualitatively leads to faster decrease of the density of survived particles. Let us also note, that our numerical and analytical results are in a proper coincidence. The best agreement between analytical and numerical estimates is obtained for the case $\delta=1$ (see Fig.~\ref{thetadelta}) similarly as it was observed in the case with ordinary diffusion \cite{vollmayrlee2018}. Therefore our numerical results support  analytical predictions of RG theory for asymptotic scaling  behaviour  of density of  surviving L\'evy flyers.

\section{Conclusions}
\label{conclusions}

In the present paper we have studied the large time behaviour of density  and correlation function of surviving particles in   two-species reaction-diffusion system described by coupled  reactions $A + A \to (0, A)$ and $A + B \to A$  where both species perform superdiffusion motion, modeled by L\'evy flights. We considered our system in a diffusion-controlled regime occurring for space dimensions $d<d_c$, where fluctuations are dominant. In this regime the scaling of density  and correlation function of surviving particles with time is characterized by nontrivial universal exponents. We have worked with the field-theoretical representation of our model obtained by mapping master equation to the field theory and neglecting terms describing ordinary diffusion.  For systems with L\'evy flights, the critical dimension is determined by control parameter of long-range hops decay: $d_c=\sigma$. Applying RG methods, we have calculated universal exponents for the scaling behaviour of density  and correlation function up to the first order in deviation from the critical dimension $\epsilon=\sigma-d$. Our analytical outcome reveals that at least in the one-loop approximations  such exponents for a case with L\'evy flights  can be obtained from expressions of their counterparts for ordinary diffusion case simply by substituting of the space dimension $d$ by an effective dimension  that scales with L\'evy flight control parameter: $d_{\rm eff}=2d/\sigma$. This is similar to the picture observed for the critical behaviour of systems with long-range interactions,  where a hypothesis was proposed: long-range critical exponents can be obtained from expressions for short-range critical exponents using instead space dimension $d$ an effective one expressed via $d$ itself and parameter of long-range interaction decay (for details see e.g.  \cite{Defenu20,Benedetti20}).

We have performed also numerical simulations of the considered process,  applying the synchronous version of cellular automaton updating algorithm. Our numerical estimates for decay exponent of surviving particles' density  corroborate  analytical predictions for this exponent. In particular, they show that probability   of target particle is higher in process with L\'evy flights comparing with process with ordinary diffusion.

\section*{Acknowledgements}

We thank Yurij Holovatch for reading the draft of the manuscript and for valuable remarks. This work was financially supported by the National Academy of Sciences of Ukraine within the framework of the Project K$\Pi$KBK 6541230.  V.B. is grateful for support from the U.S. National Academy of Sciences (NAS) and the Polish Academy of Sciences (PAS) for scientists from Ukraine. M. D. is thankful to LPTMC members for the hospitality during  stay in Sorbonne University, where his part of the work was finalized.

\clearpage

\appendix

\section{Smoluchowski approximation} \label{ap0}
As we have already mentioned in the Introduction one of the methods widely used to study fluctuation effects in different reaction-diffusion systems for dimensions below and at the critical one ($d \leq d_{c}$) is Smoluchowski theory \cite{Smoluchowski,Smoluch_appr}. The main idea of the Smoluchowski approach is to relate reaction rates to diffusion ones assuming that particles interact with each other within a fixed distance (see e. g. \cite{krapivsky1994, howard1996,rajesh2004, Oshanin1994}). For several simple reaction-diffusion models, this approach predicts the correct decay exponents (for instance, annihilation reaction \cite{Smoluchowski}), but one does not allow quantitatively calculating the amplitudes yet. 
Here we briefly discuss Smoluchowski approximation for our two-species reaction-diffusion model (\ref{model}) in the case when particles move according to  L\'evy flights. In the Smoluchowski theory, the particles may instantaneously interact with each other within a fixed distance and, thus, the reactions rates $\lambda$ and $\lambda^\prime$ are replaced by their effective counterparts related to the diffusion of the particles. Using the first return probabilities for  L\'evy flights we can get for dimensions $d < d_c = \sigma$  \begin{eqnarray}
	\lambda &=& \mbox{const}\times t^{d/\sigma - 1}, \\
	\lambda^{\prime} &=& \mbox{const}\times \left(\frac{1 + \delta}{2}\right)^{d/\sigma} t^{d/\sigma - 1}, 
\end{eqnarray}
where $\delta=D_{A}/D_{B}$ in our case with transport constants $D_A$ and $D_B$  associated with  L\'evy flights of particles.
Substituting such results into rate equations of mean densities $\langle a\rangle$ and $\langle b\rangle$ for our system
\begin{eqnarray}
	\frac{\partial \langle a\rangle}{\partial  t}=-\lambda  \langle a\rangle^2, \\
	\frac{\partial \langle b\rangle}{\partial  t}=-2Q\lambda^\prime \langle b\rangle\langle a\rangle ,
\end{eqnarray}
and solving them with respect   to   the mean density of $B$ particles we get  the following behavior for $d <  \sigma$:
\begin{eqnarray}
	\langle b \rangle_{S} \sim t^{-\theta_{S}}, \label{expTh}
\end{eqnarray}
with
\begin{eqnarray}
	\label{smoluch_exp}
	\theta_{S} =  \frac{{2} \, d}{\sigma (2 - p)} \left(\frac{1 + \delta}{2}\right)^{d/\sigma}.
\end{eqnarray}

Despite the acceptability and convenience of this theory, such an approach may predict erroneous results for a more complex system. In particular, Smoluchowski theory does not allow obtaining the correct exponents for the mixed reactions (for more details, see \cite{tauber2005}). Moreover, one gives incorrect results for well-known exactly solvable models (see e. g. \cite{krishnamurthy}).

\section{One Loop Contributions} \label{apI}
\subsection{Contributions for density $\langle b \rangle$} \label{ap1}

\begin{figure}
	\begin{center}
		\includegraphics[width=110mm]{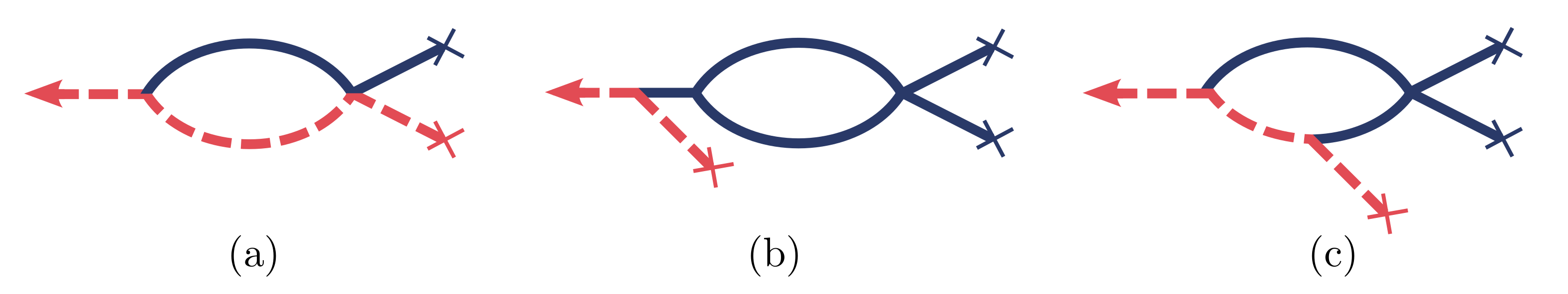}
		\caption{Diagrams presenting one-loop contributions  to the mean-field density 	$\langle b(t)\rangle$ 
		}
		\label{one-loop_density}	
	\end{center}
\end{figure}
Here we present results for one-loop contributions to  density  of $B$ particles. Diagram representations are given in Fig.~\ref{one-loop_density}. Results of calculation for these diagrams in the limit $a_0\to \infty$ are follows
\begin{eqnarray}
	\label{contra}
	\mbox{(a)}&=& \frac{{\lambda^\prime}^2 Q b_{0}}{\lambda(a_{0} \lambda t)^{Q \lambda^{\prime} / \lambda}} \frac{\Gamma{(d/\sigma)}}{(4\pi)^{d/2}\Gamma{(d/2)}} t^{\epsilon/\sigma} \frac{2}{(1 + \delta)^{d/\sigma}} \frac{\sigma^{2}}{\epsilon^{2} (\epsilon + \sigma)}    \\ 
	\label{contrb}
	\mbox{(b)}&=&  -\frac{{\lambda^\prime} Q b_{0}}{(a_{0} \lambda t)^{Q \lambda^{\prime} / \lambda}} \frac{\Gamma{(d/\sigma)}}{(4\pi)^{d/2}\Gamma{(d/2)}} \lambda^{\prime} t^{\epsilon/\sigma} \frac{8 }{2^{d/\sigma}} \frac{\sigma^{4}}{\epsilon^{2} (\epsilon + \sigma)^{2} (\epsilon + 2 \sigma)} \\
	\label{contrc}
	\mbox{(c)}&=&  \frac{{\lambda^\prime}^2 Q^2 b_{0}}{\lambda(a_{0} \lambda t)^{Q \lambda^{\prime} / \lambda}}  \frac{\Gamma{(d/\sigma)}}{(4\pi)^{d/2}\Gamma{(d/2)}} \lambda^{\prime} t^{\epsilon/\sigma} \frac{1 }{(1 + \delta)^{d/\sigma - 1}} \left(\frac{f(\delta)}{\epsilon}  + O(\epsilon)\right),
\end{eqnarray} 
with $f(\delta)$ given by (\ref{f(delta)}).
Results for the first two diagrams were obtained in \cite{birnsteinova2020}, for the third diagram we have calculated the leading contribution at $\epsilon\to 0$.

\subsection{Contributions for correlation function ${C}_{BB}$} \label{ap2}

Here we present the contributions to $\hat{C}_{BB}( k = 0)$ at one loop.
All possible diagrams of six topology classes are presented schematically in Fig.~\ref{one-loop_corr}  (see also \cite{vollmayrlee2018}). Results of the calculation of poles in $\epsilon = \sigma - d$ can be presented in the following form:
\begin{eqnarray}
	\hat{C}_{BB}^{1-loop}(k = 0) = \frac{1}{3(4 \pi)^{d/2}} \frac{\Gamma(d/\sigma)}{\Gamma(d/2)} \frac{b_{0}^{2} Q \lambda^{\prime \, 2} }{(a_{0} \lambda t)^{2 Q \lambda^{\prime}/ \lambda}} t^{1 + \epsilon/\sigma}\sum_{i = 1}^{6} F_{i} (z, \epsilon, \sigma),
\end{eqnarray}
where $F_{i}$ are all possible contributions to $\hat{C}_{BB}( k = 0)$ from the diagrams of six topology classes with shortening $z = \lambda^{\prime} / \lambda$:
\begin{eqnarray}
	\label{F1}
	F_{1} &=& \left(4Q - \frac{6z}{1 + \delta}\right)\frac{1}{\epsilon}, \\
	\label{F2}
	F_{2} &=& 6 \sigma Q z \left(\frac{z}{1 + \delta} - 1\right)\frac{1}{\epsilon^{2}} \nonumber \\
	&+& \left(6 - 8 Q + 24 Q z + 3 Q^{2} z^{2} f\left(\delta\right) - 15 \frac{Q z^{2}}{1 + \delta}\right)\frac{1}{\epsilon}, \\
	\label{F3}
	F_{3} &=& \left(6 - \frac{14}{3}Q + \frac{-6 z + 8 Q z + 15 Q z^{2} - 10 Q^{2} z^{2}}{1 + \delta}\right)\frac{1}{\epsilon}, \\
	\label{F4}
	F_{4} &=& 6 \sigma Q z \left(1 - \frac{4 Q}{3}\right)\left(\frac{z}{1 + \delta} - 1\right) \frac{1}{\epsilon^{2}} \nonumber \\
	&+& \left(-6 + \frac{28 Q}{3} + 6 Q z - 20 Q^{2} z + \left(3 Q^{2} z^{2} - 4 Q^{3} z^{2}\right) f\left(\delta\right) \right. \nonumber \\ 
	&+& \left. \frac{6}{1 + \delta}\left(3 Q^{2} z^{2} - 2 Q z^{2}\right)\right)\frac{1}{\epsilon},
\end{eqnarray}
with $f(\delta)$ defined in Eq. (\ref{f(delta)}) and $F_5=0$ and $F_6=0$ since there is no poles in $\epsilon$ for these diagram classes.

\begin{figure}
	\begin{center}
		\includegraphics[width=100mm]{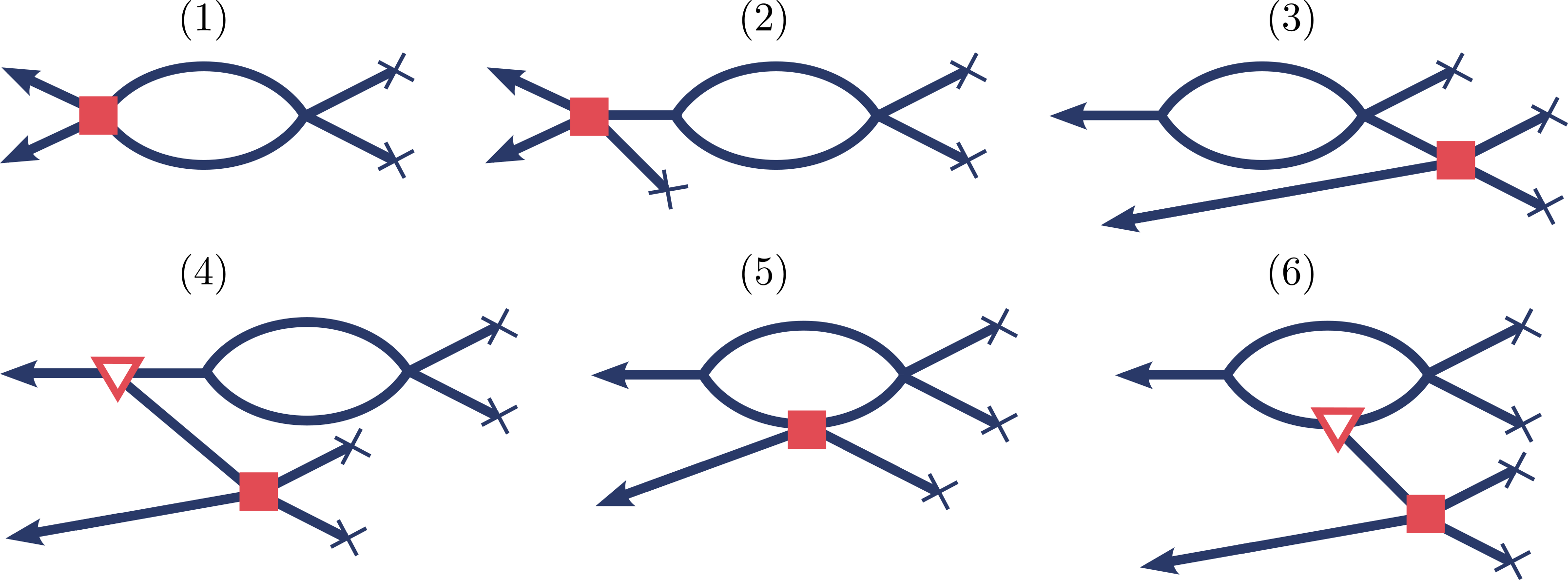}
		\caption{One-loop corrections  to the correlation function with six topology classes of the diagrams (without distinguishing between $A$ and $B$ lines), where three-point vertex are shown by empty triangles, while four-point vertices  are shown by filled rectangles. For more details see \cite{vollmayrlee2018}}
		\label{one-loop_corr}	
	\end{center}
\end{figure}



\end{document}